\documentclass[11pt]{article}

\usepackage[dvips]{graphics}
\def\ra{\rightarrow}
\def\iy{\infty}

\def\be{\begin{equation}}
\def\ee{\end{equation}}
\def\ba{\begin{eqnarray*}}
\def\ea{\end{eqnarray*}}
\def\bae{\begin{eqnarray}}
\def\eae{\end{eqnarray}}
\def\bc{\begin{center}}
\def\ec{\end{center}}
\def\ov{\over}

\def\s{\sigma}

\def\la{\lambda}
\def\pr{\textrm{Prob}}

\def\th{\theta}

\begin{document}
\title{Universality of the Distribution Functions \\
of\\
 Random Matrix Theory}
\date{August 31, 1999}
\author{
Craig A.~Tracy\\
Department of Mathematics\\
Institute of Theoretical Dynamics\\
University of California\\
Davis, CA 95616, USA
\and
Harold Widom\\
Department of Mathematics\\
University of California\\
Santa Cruz, CA 95064, USA
}
\maketitle
\section{Random Matrix Models}
\setcounter{equation}{0}
In  probability theory and statistics  a common first approximation
to many random processes is
a sequence 
$X_1, X_2, X_3, \ldots $
of independent and identically distributed (iid) random variables.  Let $F$
denote their common distribution.  To motivate the material below,
we take  these random variables and construct 
a particularly simple $N\times N$ random matrix,
\[ \textrm{diag}\left(X_1(\omega),X_2(\omega),\cdots,
 X_N(\omega)\right). \]  
The order statistics are  the eigenvalues ordered
\[ \la_1\le \la_2\le \cdots \le \la_N, \]
and the distribution of the largest eigenvalue, 
$\la_{\textrm{max}}(N)=\la_N$, is
\ba
\pr\left(\la_{\textrm{max}}(N)\le x\right)&=& \pr\left(X_1\le x,\ldots, X_N\le x\right)\\
 &=& F(x)^N.
\ea
Since the distribution $F$ is arbitrary, we observe that so too is the distribution of the largest
eigenvalue of a $N\times N$ random matrix. However, one is really interested in limiting
laws as $N\ra\iy$.  That is, we ask if there exist constants $a_N$ and $b_N$ such that
\be {\la_{\textrm{max}}(N)-a_N \over b_N} \label{normEig}\ee
converges in distribution to a nontrivial limiting distribution function $G$.
In the present situation a complete answer is provided by

\textit{Theorem:} If (\ref{normEig}) converges in distribution
to some nontrivial distribution function $G$, then $G$ belongs to one of the following
forms:
\begin{enumerate}
\item $e^{-e^{-x}}$ with support \textbf{R}.
\item $e^{-1/x^\alpha}$ with support $[0,\iy)$ and $\alpha>0$.
\item $e^{-(-x)^\alpha}$ with support $(-\iy,0]$ and $\alpha>0$.
\end{enumerate}

This theorem is a model for the type of results we
want for  nondiagonal random matrices.

A random matrix model is a probability space $(\Omega,\mathcal{P},\mathcal{F})$ where
$\Omega$ is a set of matrices.  Here are some examples

\begin{itemize}
\item Circular Unitary Ensemble (CUE, $\beta=2$)
\begin{itemize}
\item $\Omega=\mathcal{U}(N)=N\times N$ unitary matrices.
\item $\mathcal{P}$= Haar measure.
\end{itemize}
\item Gaussian Orthogonal Ensemble (GOE, $\beta=1$)
\begin{itemize}
\item $\Omega=N\times N$ real symmetric matrices.
\item $\mathcal{P}$ = unique\footnote{Uniqueness is up
to centering and a normalization of the variance.} measure that is  invariant
under orthogonal  transformations and the  matrix elements
(say on and above the diagonal) are iid random
variables.
\end{itemize}
\item Gaussian Unitary Ensemble (GUE, $\beta=2$)
\begin{itemize}
\item $\Omega= N\times N$ hermitian matrices.
\item $\mathcal{P}$= unique measure that is invariant
under unitary transformations and the real and imaginary matrix
elements (say on and above the diagonal) are iid random variables.
\end{itemize}
\item Gaussian Symplectic Ensemble (GSE, $\beta=4$)
\begin{itemize}
\item $\Omega=2N\times 2N$  Hermitian self-dual
matrices.\footnote{Identify the $2N\times 2N$ matrix with 
the $N\times N$ matrix whose entries are quaternions.  If the
quaternion matrix elements satisfy ${\overline M_{ji}}=M_{ij}$ where
the bar is quaternion conjugation, then the $2N\times 2N$ matrix
is called Hermitian self-dual.  Each eigenvalue of a Hermitian
self-dual matrix has multiplicity two.}
\item $\mathcal{P}$= unique measure that is invariant
under symplectic transformations and the real and imaginary matrix
elements (say on and above the diagonal) are iid random variables.
\end{itemize}
\end{itemize}
 
Expected values of random variables
$f:\Omega\ra\textbf{C}$ are computed from the usual
formula
\[ E_\Omega(f) = \int_\Omega f(M) \, d\mathcal{P}(M). \]
If $f(M)$ depends only on
the eigenvalues of $M\in\Omega$, then one can
be more explicit:
\begin{itemize}
\item CUE (Weyl's Formula)
\[
E_{\mathcal{U}(N)}(f) ={1\over N! (2\pi)^N}
\int_{-\pi}^\pi\cdots \int_{-\pi}^\pi
f(\theta_1,\ldots,\theta_N)
\prod_{1\le\mu<\nu\le N}
\left| \Delta(e^{i\th_1},\ldots,e^{i\th_N})\right|^2
\, d\theta_1\cdots d\theta_N , \]
\item Gaussian Ensembles ($\beta=1,2,4$):
\end{itemize}
\[
E_{N\beta}(f)= c_{N\beta}
 \int_{-\iy}^\iy\cdots\int_{-\iy}^\iy
f(x_1,\ldots,x_N)
\left\vert\Delta(x_1,\ldots,x_N)\right\vert^{\beta}
e^{-{\beta\over 2}\sum x_j^2}
\, dx_1\cdots dx_N , \]
where $c_{N\beta}$ is chosen so that $E_{N\beta}(1)=1$
and $\Delta(x_1,\ldots,x_N)=\prod_{1\le i<j\le N}(x_i-x_j)$.
The factor $e^{-{\beta\over 2}\sum x_j^2}$ explains the choice of the
word ``gaussian'' in the names of these ensembles.
A commonly studied generalization of these gaussian measures is
to replace the sum of quadratic terms appearing in the exponential
with $\sum V(x_i)$ where $V$ is, say, a polynomial (with the obvious restrictions
to make the measure well-defined).     
 
Choosing $f=\prod_i\left(1-\chi_{\raisebox{-.5ex}{$\scriptstyle{J}$}}(x_i)\right)$, 
$\chi_J$  the characteristic function of a set $J\subset \textbf{R}$, we
get the important quantity\footnote{This quantity has an obvious extension to
other random matrix models.} 
\[ E_{N\beta}(f)=E_{N\beta}(0;J):=\textrm{probability no eigenvalues lie in}\> J,\]
and in the particular case $J=(t,\iy)$
 \[ F_{N\beta}(t):=\textrm{Prob}\left
(\la_{\textrm{max}}\le t\right)=E_{N\beta}(0,J).\]
The level spacing distribution\footnote{Let the eigenvalues be ordered.  The conditional
probability that given an eigenvalue at $a$, the next one lies between $s$ and $s + ds$
is called the level spacing density.} is expressible in terms of the mixed second partial
derivative of $E_{N\beta}(0;(a,b))$ with respect to the endpoints $a$ and $b$.

\section{Fredholm Determinant Representations}
\setcounter{equation}{0}
Though $E_{N\beta}(0;J)$ are explicit  $N$-dimensional integrals, these expressions
are not so useful in establishing limiting laws as $N\ra\iy$.  What turned out to be
very useful are Fredholm determinant representations for $E_{N\beta}(0;J)$.  In 1961
M.~Gaudin proved for $\beta=2$ (using the newly developed
orthogonal polynomial method of M.~L.~Mehta) 
 that $E_{N2}(0;J)=\det(I-K_{N2})$ where $K_{N2}$
is an integral operator acting on $J$ whose kernel is of the form
\be  {\varphi(x)\psi(y)-\psi(x)\varphi(y)\over x-y }\, ,\label{kernel}\ee
with
$\varphi(x) =c_N e^{-x^2/2} H_N(x)$,  
$\psi(x) =c_N e^{-x^2/2} H_{N-1}(x)$,
and $H_j(x)$ are the Hermite polynomials.\footnote{For the random matrix models
corresponding to general potential $V$,  $\varphi(x) =c_N e^{-V(x)/2} p_N(x)$
and $\psi(x)=c_N e^{-V(x)/2} p_{N-1}(x)$ where $p_j(x)$ are the orthogonal
polynomials associated with weight function $w(x)=e^{-V(x)}$.  It is in this generalization
that we see the close relation between the general theory of orthogonal polynomials
and random matrix theory.}  For $\beta=1\>\textrm{or}\> 4$, generalizing F.~J.~Dyson's 1970 analysis
of the $n$-point correlations for the circular ensembles, it follows from
work by Mehta the following
year that the square of $E_{N\beta}(0;J)$ again equals a Fredholm determinant,
$\det(I-K_{N\beta})$, but now the kernel of $K_{N\beta}$ is a $2\times 2$ matrix.\footnote{
See \cite{tw1} for  elementary proofs of these facts.}
\section{Scaling Limits (Limiting Laws)}
\setcounter{equation}{0}
\subsection{Bulk Scaling Limit}
Let $\rho_N(x)$ denote the density of eigenvalues at $x$ and
pick a point $x_0$, independent of $N$ with $\rho_N(x_0)>0$. We scale distances
so that resulting density is one at $x_0$,
$\xi:=\rho_N(x_0)\left(x-x_0\right)$,
and we call the limit
\[ N\rightarrow\infty, x\rightarrow x_0,\> \textrm{such that}\>\>\xi
\>\> \textrm{is fixed},\]
the bulk scaling limit.  For $\beta=2$,
\[ E_{N 2}(0;J)\rightarrow E_2(0;J)
=\det\left(I-K_2\right) \]
where the integral operator $K_2$ (acting on $L^2(J)$) has
as its kernel (the sine kernel) 
\[ {1\over \pi}
{\sin \pi(\xi-\xi^\prime)\over \xi -\xi^{\prime}}\, .\]
(We use the same symbol $J$ to denote the scaled
set $J$.)
Furthermore, 
\[ p_2(s)=-{d^2\ov ds^2} E_{2}\left(0; (0,s)\right)\]
is the (limiting) level-spacing density for GUE; known as
the Gaudin distribution.\footnote{For the 
analogous $\beta=1,4$ results, see, e.g.,
\cite{mehta} or \cite{tw1}.}
We observe that the limiting kernel
is translationally invariant and  independent of $x_0$.

\subsection{Edge Scaling Limit}
In the gaussian ensembles, the density decays
   exponentially fast around
$2\sigma\sqrt{N}$; perhaps surprisingly, it is also the case that 
\be \lim_{N\rightarrow\infty}{\lambda_{\textrm{max}}(N)\over \sqrt{N}}
= 2\s,\>
a.s.\label{approxFirst}\ee
where $\s$ is the standard deviation of the off-diagonal
matrix elements.  (In the normalization we've adopted, $\s=1/\sqrt{2}$.)
If we introduce the scaled random
variable $\hat\la$ through 
\[ \la_{\textrm{max}} = 2\s\sqrt{N} +
 {\sigma {\hat\la}\over N^{1/6}}\, , \]
then
\[\textrm{Prob}\left(\la_{\textrm{max}}\le t\right)
=\textrm{Prob}\left({\hat\la}\le s\right)\rightarrow F_\beta(s) \>\> \textrm{as}\>\> N\ra\iy,\]
where
$ t=2\s\sqrt{N}+\s s/N^{1/6}$. For $\beta=2$, 
\[F_2(s)=\det(I-K_{\textrm{Airy}}),\]
where $K_{\textrm{Airy}}$ has kernel of the form (\ref{kernel}) with
$\varphi(x)= \textrm{Ai}(x)$,   $\psi(x)=\textrm{Ai}^\prime(x)$
and $J=(s,\iy)$.
(See, e.g., \cite{tw1} for the $\beta=1,4$ results.)

\section{Connections with Integrable Systems}
\setcounter{equation}{0}
\subsection{Bulk Scaling Limit}
In 1980 M.~Jimbo, T.~Miwa, Y.~M\^ori, and M.~Sato~\cite{jmms}
expressed the Fredholm determinant of the sine kernel
 in terms of a solution to
a certain system of integrable differential
equations.\footnote{A simplified proof of
their results can be found in \cite{tw2}.}  In the simplest case of a single interval, 
$J=(0,s)$, the differential equation is a particular
case of Painlev\'e V ($P_V$)\footnote{The
differential equation below is the sigma representation of $P_V$.}
 and the Fredholm determinant is given by
\[
\det\left(I-\lambda K_2\right)=
\exp\left(\int_0^{\pi s} {\s(x;\lambda)\over x}\,dx\right),\]
\[
\left(x\s^{\prime\prime}\right)^2
+4\left(x\s^\prime-\s\right)\left(x\s^\prime
-\s+(\s^\prime)^2\right)=0,\]
\[ \s(x;\la)\sim -{\la\over\pi}\, x,\>\>\textrm{as}\>\> x\rightarrow
0.\]
For $\beta=1,4$ and $J=(0,s)$, $E_\beta(0;(0,s))$ can also be
expressed in terms of the same function $\s(x;1)$.  A down-to-earth
application of these Painlev\'e representations (and
using the known asymtptotics of $\s(x;1)$) is that one can
easily produce graphs of the level spacing densities $p_\beta(s)$.\footnote{Without
the Painlev\'e representations, the numerical evaluation of the Fredholm
determinants is quite involved.}
\subsection{Edge Scaling Limit}
The limiting distributions (edge scaling) of the largest eigenvalue,  $F_\beta(s)$, can
also be expressed in terms of Painlev\'e functions---this time $P_{II}$ \cite{tw3,tw4}:
\bae
F_1(s)^2 & = & \exp\left(-\int_s^\iy q(x)\, dx\right)\, F_2(s), \label{F1}\\	
F_2(s) &=&\exp\left(-\int_s^\iy (x-s) q(x)^2\, dx \right), \label{F2}\\
F_4(s/\sqrt{2})^2&=& \cosh^2\left({1\ov 2}\int_s^\iy q(x)\, dx\right) \, F_2(s),
\label{F4}\eae
where $q$ satisfies the  Painlev\'e II equation
\[ q^{\prime\prime}=x q + 2 q^3\]
with boundary condition $q(x)\sim \textrm{Ai}(x)$ as $x\ra\iy$.\footnote{That
a solution $q$ exists and is unique follows from the representation
of the Fredholm determinant in terms of it.  Independent proofs 
of this,  as well as the asymptotics as $x\ra -\iy$,  were
given by S.~Hastings and J.~McLeod,  
P.~Clarkson and McLeod and by P.~Deift and X.~Zhou.}  
The graphs of
the densities $f_\beta(s)=dF_\beta(s)/ds$ are in Figure 1.
\begin{figure}
\vspace{-2.5cm}
\hspace{2cm}\resizebox{10cm}{10cm}{\includegraphics{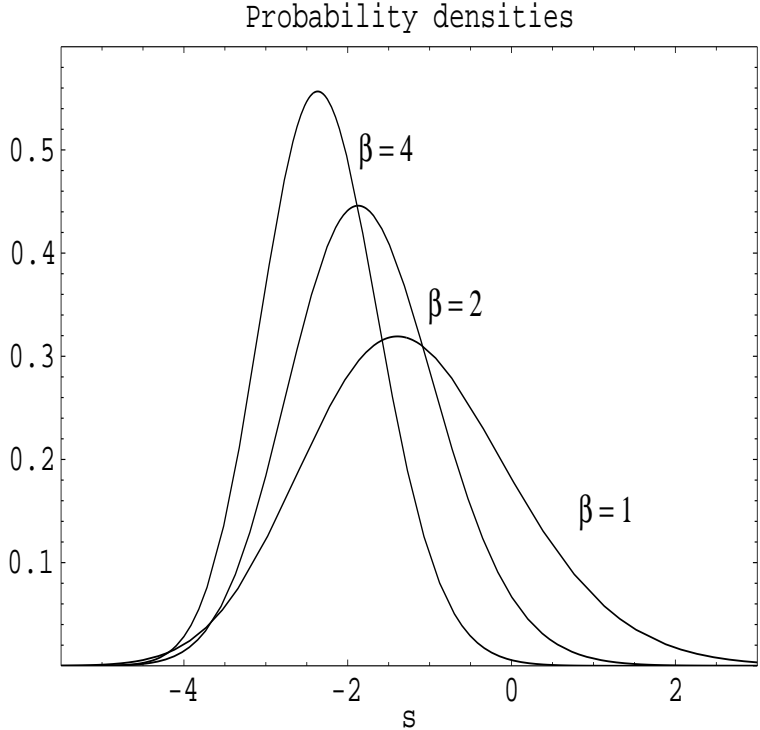}}	 
\caption{Densities for the scaled largest eigenvalues, $f_\beta(s)$.}
\end{figure}
\subsection{Generalizations}
Both the sine kernel and the Airy kernel are of the form (\ref{kernel}).
Kernels of this form arise in many problems in integrable systems; indeed,
so much so that A.~Its, A.~Izergin, V.~Korepin and V.~Slavnov~\cite{its} in 1990 initiated
a general analysis of these kernels.  The following theorem \cite{tw5}, which
applies to a wide class of $\beta=2$ random matrix ensembles, 
gives the general situation:
 
\textit{Theorem:} Let $J=\bigcup_{j=1}^m(a_{2j-1}-a_{2j})$ be a union of
 open intervals. Define
 $\tau(a)=\det\left(I-K\right) $
 where $K$ is an integral operator acting on $L^2(J)$ whose
 kernel is of the form (\ref{kernel})
 where $\varphi$ and $\psi$ are assumed to satisfy
 \[ {d\over dx}
 \left(
 \begin{array}{c} \varphi\\ \psi \end{array}
 \right)= \Omega(x)\left(\begin{array}{c} \varphi\\ \psi \end{array}\right)
 \]
 with $\Omega(x)$ a $2\times 2$ matrix with zero trace and
 rational entries in $x$,
  then
 $ {\partial \over \partial a_j}\log\det(I-K) $
  are expressible polynomially in terms of solutions to a total
  system of partial differential equations ($a_j$ are the 
  independent variables).  The differential equations are given
  explicitly in terms of the coefficients of the rational functions
  appearing in $\Omega(x)$.
  
\subsection{Historical Comments}
The first connection between  Toeplitz/Fredholm determinants
and Painlev\'e functions was established in 1973--77 in work of  T.~T.~Wu,
B.~M.~McCoy, E.~Barouch and the first author concerning
the scaling limit of the 2-point functions of the 2D Ising
model of statistical mechanics.   The Painlev\'e function
that arose was $P_{III}$.
This work was subsequently generalized by Sato, Miwa and Jimbo to 
$n$-point functions and, more
generally,  holonomic quantum fields.   The Kyoto School then
took up the problem of the density matrix of the impenetrable Bose
gas and it was in this context that they discovered that the Fredholm determinant
of the sine kernel is related to $P_V$.  

A crucial simplification of the Kyoto School
work, as it applies to random matrix theory, was made by Mehta in 1992~\cite{mehta2}.  This last
work inspired the commutator methods introduced by the present authors in the period 1993--96.
Since then both Riemann-Hilbert methods of Deift, Its, Zhou  and others (see, e.g.~\cite{diz}); and Virasoro
methods  of M.~Adler, T.~Shiota, P.~van Moerbeke,  and others (see, e.g.~\cite{atm}),  
have played an increasingly important
role in the development of  random matrix theory. The connection of these methods with
the isomonodromy method has been clarified by J.~Palmer~\cite{palmer} and J.~Harnad~\cite{harnad}.

Space does not permit us to discuss the interesting
  connections between random matrices and
Szeg\"o type limit theorems.   See E.~Basor~\cite{basor1} for connections
with linear statistics and
the  review papers \cite{basor2, tw6} for some related historical comments.
\newpage
\section{Universality}
\setcounter{equation}{0}

\subsection{Universality of Gaussian Ensembles in Random Matrix Models}
\subsubsection{Invariant Measures, $\beta=2$}
As briefly mentioned above, a widely studied class of random matrix
models is defined by the replacement of the gaussian potential, $x^2$,
by the general potential $V(x)$.  For the weight functions most studied,
the parameter $N$ is put into the exponent so that the weight function
becomes $e^{-N V(x)}$.  
For different $V$'s, the limiting density $\rho_V(x)$ can be quite different.
It may be supported on many distinct intervals, and it may vanish
at interior points of its support.  In the gaussian case, the limiting
density is the Wigner semicircle law: $\rho_W(x)={2\ov\pi}\sqrt{1-x^2}$.
Heuristic arguments suggest that the behavior exhibited by the
Wigner law---that $\rho$ is positive on the interior of its
support and vanishes like a square root at endpoints---is the
typical behavior for $\rho_{V}$.  The bulk scaling limit and
edge scaling limit are defined in analogous ways to the gaussian
cases.  To establish universality of these scaling limits, one must
show (for $\beta=2$ ensembles) that the scaled kernels approach
the sine kernel and the Airy kernel, respectively. 
The potential $V(x)={t\ov 2} x^2+{g\ov 4} x^4$ ($g>0$, $t<0$) is an
example of a ``two interval'' potential.  Indeed, for this important 
example P.~Bleher and A.~Its~\cite{bleher} 
proved precisely this statement of universality.  (See their paper for 
 related work in the orthogonal polynomial literature
as well as the physics literature.) Recently, building on
work of \cite{dkmvz}, 
A.~Kuijlaars and
K.~McLaughlin \cite{ken} have shown this behavior is generic for real analytic $V$
satisfying $\lim_{x\ra\iy} V(x)/\log\vert x\vert = +\iy$.

In the physics literature,
M.~Bowick, E.~Br\'ezin~\cite{bowick} and others have argued (for $\beta=2$ 
ensembles) that if 
$\rho_V$ vanishes faster than a square root, then the corresponding
edge scaling limit will result in   nonAiry universality classes.  
The resulting new kernels will  have
form (\ref{kernel}) and the theory developed in \cite{tw5} will 
apply, but there remains much to be understood in these cases.

For $\beta=1, 4$, the situation is more complicated due to
the structure of $K_V$~\cite{tw1,w1},  and the ``universality'' theorems
are not so general.
 
\subsection{Noninvariant Measures: Wigner Ensemble}
The Wigner ensembles are defined by
requiring that the matrix elements on and above
the diagonal in either the real symmetric
case or the complex hermitian case are independent and identically
distributed random variables.  It is only in the case when the
distribution is gaussian is the measure invariant.
One usually assumes, as we do here, that all moments of the common
distribution function exist.  It was Wigner himself who showed that the
limiting density of states is the Wigner semicircle.  Subsequently
several authors---culminating in a theorem
by Z.~Bai and Y.~Lin clarifying which
moments need exist---showed (\ref{approxFirst}) continues to hold
for the Wigner ensembles.  

It should be noted that because the measure is noninvariant, the
nongaussian Wigner ensembles do not, as far as we understand, have
Fredholm determinant representation for their distribution functions.
This means, for one, that the methods of integrable systems are not
directly applicable to Wigner ensembles.
It is therefore particularly important,  as  A.~Soshnikov \cite{soshnikov}
recently proved, that in the edge scaling limit  the Wigner ensembles are  
in the same universality class
as the gaussian models.  In particular, the limiting distribution of the scaled largest
eigenvalue is given by $F_1(s)$ for real symmetric Wigner matrices and by $F_2(s)$
for complex hermitian Wigner matrices.

\subsection{Examples from Physics}
A second type of universality, and the one first envisioned by Wigner
in the context of nuclear physics,  asserts in Wigner's words \cite{wigner}
\begin{quote}
Let me say only one more word.  It is very likely that the curve in
Figure I [an approximate graph of $p_1(s)$] is a universal function.  In other
words, it doesn't depend on the details of the model with which
you are working.
\end{quote}
The modern version of this asserts 
that for a classical, ``fully'' chaotic Hamiltonian the corresponding quantum
system has a level spacing distribution equal to $p_\beta(s)$ in the bulk.  (The
symmetry class determines which ensemble.)  
This quantum chaos conjecture, due  to O.~Bohigas, M.~Giannoni and C.~Schmit \cite{bohigas},
has been a guiding principle for much subsequent work, though it is the authors' understanding that
it remains a conjecture.  
A particularly nice numerical example supporting
this conjecture is M.~Robnik's work \cite{robnik} on chaotic billards.
The reader is referred to the recent review article \cite{guhr} for further numerical examples
that support this conjecture.  It should  be noted that there are examples from
number theory where the conjecture fails.  Thus, as it has been said, the conjecture is
undoubtedly true except where it is demonstratively false.  

\subsubsection{Aperiodic Tiling Adjacency Matrix}
The discovery of quasicrystals has made the study of statistical
 mechanical models whose underlying lattice is quasiperiodic of considerable
 interest to physicists.  In particular,  in order to understand
 transport properties, tight binding models have been defined
 on various quasiperiodic lattices.  One such study
 by Zhong \textit{et al.}~\cite{grimm}  defined a simplified tight binding model
 for the octagonal tiling of Ammann and Beenker.  This quasiperiodic
 tiling consists of squares and rhombi with all 
 edges of  equal lengths (see Figure 2)
  and has a $D_8$ symmetry around the central vertex.
 On this tiling the authors take as their Hamiltonian the adjacency
 matrix for the graph with free boundary conditions.  The largest
 lattice they consider has 157,369 vertices.  The matrix splits
 into ten blocks according to the irreducible representations
 of the dihedral group $D_8$.  For each of these ten independent subspectra,
 they compare the empirical 
 distribution of the normalized   spacings of the
 consecutive eigenvalues with the
 GOE level spacing density
 $ p_1(s)$.  In Figure 2 we have
 reproduced a  portion of their data for one
 such subspectrum together with  $p_1$.  
  \begin{figure}
\begin{center}
\vspace{3.2cm}
\resizebox{11cm}{11cm}{\includegraphics{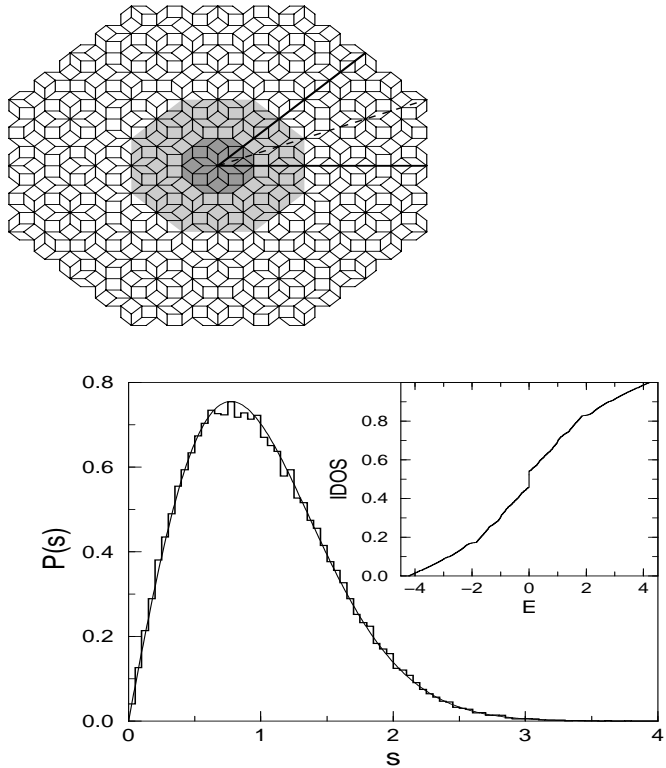}}
\end{center}
\vspace{-7cm}
\caption{Data for nearest neighbor
normalized spacings of eigenvalues
of the adjacency matrix for a quasiperiodic octagonal tiling 
 are plotted together with the GOE level spacing  density $p_1(s)$.  
Data are  from one independent subspectrum of a $D_8$-symmetric
octagonal patch of a tiling with 157,369 vertices. Courtesy of
Zhong \textit{et al.}~\cite{grimm}.}
\end{figure}

\subsection{Spacings of the Consecutive Zeros of Zeta Functions}
Perhaps the most surprising appearance of the distributions of random
matrix theory is in number theory.  Analytical work by H.~Montgomery
and extensive numerical calculations by A.~Odlyzko on the zeros
of the Riemann zeta function have given convincing evidence that the normalized
consecutive spacings follow the Gaudin distribution, see Figure 3. 
 Recent results of Z.~Rudnick and P.~Sarnak are also compatible with the belief
 that the distribution of the spacings between zeros, not only of
 the Riemann zeta function, but also of quite general automorphic
 $L$-functions over \textbf{Q}, are all given by this Montgomery-Odlyzko Law.
 In their landmark book \cite{katz}, N.~Katz and P.~Sarnak establish the
 Montgomery-Odlyzko Law for wide classes of zeta and $L$-functions
 over finite fields.  
 \begin{figure}
 \vspace{-1.5cm}
 \begin{center}
 \resizebox{11cm}{11cm}{\includegraphics{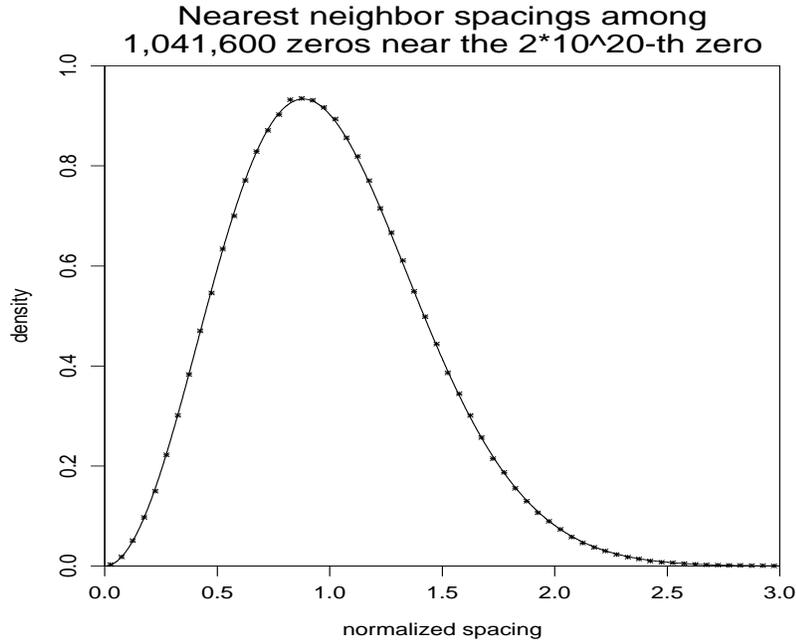}}
 \end{center}
 \vspace{-2cm}
 \caption{Data for nearest neighbor spacings among 1,041,600 zeros
of the Riemann zeta function near the $2\times 10^{20}$-th zero
are plotted together with the GUE spacing density.  Courtesy of
Andrew Odlyzko~\cite{odlyzko}.}
\end{figure}

\subsection{Random Matrix Theory and Combinatorics}
The last decade has seen a flurry of activity centering
around connections between combinatorial probability
of the Robinson-Schensted-Knuth (RSK) type on
the one hand and random matrices and integrable systems
on the other.  From the point of view of probability theory,
the quite surprising feature of these developments is that
the methods came from Toeplitz determinants, integrable
differential equations of the Painlev\'e type and the closely
related Riemann-Hilbert techniques as they were applied
and refined in random matrix theory.  Using these techniques
new, and apparently quite universal, limiting laws have
been discovered.  The earliest signs of these connections
can be found in the work of A.~Regev \cite{regev}
 and I.~Gessel \cite{gessel}.  Here, however, we introduce
 this subject by examining a certain card game
 of D.~Aldous and P.~Diaconis \cite{aldous2}, called
 patience sorting. 
\subsubsection{Patience Sorting and Random Permutations}
Our deck of cards is labeled  $\{1,2,\ldots,N\}$ and we
order the cards with their natural ordering.  Shuffle
the deck of cards and
\begin{itemize}
\item Turn over the first card.
\item Turn over the second card. If it is of higher rank,
start a new pile to the right of the first card.  Otherwise place
the second card on top of the first card.
\item Turn over the third card.  If it is of higher rank than
either the first or the second card, start a new pile to the right of
the second card.  Otherwise place  the third card on top of the card
of higher rank.  If both first and second are of higher rank, place
the third card on the smaller ranked card.  (That is, play cards as far
as possible to the left.)
\item Continue playing the game, playing
cards as far left as possible, until all the cards are turned over.
\end{itemize}
 The object of the game is to end with a small number of piles.
 Let $ \ell_N(\s) $
equal the number of piles at the end of the game where we started
with  deck $\s=\{i_1,i_2,\ldots,i_N\} $.
Clearly, $1\le \ell_N(\s)\le N$, but what are some typical
values for a shuffled deck?
Starting each time with
a newly shuffled deck of $N=52$ cards,  the computer played patience sorting
100,000 times.  Here are the statistics for $\ell_{52}$:  
\begin{itemize}
\item Mean=11.56 (11.00).
\item Standard Deviation=1.37 (1.74)
\item Skewness=0.33 (0.22)
\item Kurtosis Excess =0.16 (0.09)
\item Sample Range = 7 to 19 (Probability 0.993)
\end{itemize}
where the numbers in parentheses are the asymptotic predictions
(as the number of cards tends to infinity) 
of the theory of J.~Baik, P.~Deift and K.~Johansson \cite{bdj1} to
be described below.

A shuffled deck of cards,
$ \s=\{i_1,i_2,\ldots,i_N\} $,
is a permutation of $\{1,2,\ldots,n\}$, and so we think
of the shuffled deck as a random permutation.
A moment's reflection will convince
the reader that $\ell_N(\s)$  is  equal to the length of the longest
increasing subsequence in the permutation $\s$.
As a  problem in random permutations,  
determining the asymptotics of $E(\ell_N)$ as $N\ra\iy$   is called
Ulam's Problem.  In the 1970's A.~Vershik  and  S.~Kerov and independently
B.~Logan and  L.~Shepp showed
$ E(\ell_N)\sim 2\sqrt{N}$
with important earlier work by J.~Hammersley.  Hammersley's analysis
introduced a certain interacting particle system interpretation.  This was
developed by Aldous  and Diaconis~\cite{aldous1} who in 1995  gave a ``soft'' proof of
this result using
hydrodynamic scaling arguments from interacting particle theory.

Introducing the exponential generating function
\[ \sum_{N\ge 0} \textrm{Prob}(\ell_N\le n)\, {t^{N}\ov N!}\, , \]
Gessel  showed that it is equal to $D_n(t)$,  
the determinant of the $n\times n$ Toeplitz determinant with
symbol $e^{\sqrt{t}(z+z^{-1})}$.  (Recall that the $i,j$ entry
of a Toeplitz matrix equals the $i-j$ Fourier coefficient
of its symbol.)  It is in this work of Gessel and subsequent
work of Odlyzko \textit{et al.}~\cite{odlyzko2} and E.~Rains~\cite{rains}, that the
methods of random matrix theory first appear in RSK
type problems.\footnote{Gessel~\cite{gessel} does not mention
random matrices, but in light of well-known formulas in random matrix
theory relating Toeplitz determinants to expectations over the unitary
group, we believe it is fair to say that the connection with random
matrix theory begins with this discovery.  See, however, Regev~\cite{regev}.}

Starting with this Toeplitz determinant representation, Baik, Deift
and Johansson~\cite{bdj1}, using the steepest descent method
for Riemann-Hilbert problems~\cite{dz}, derived a delicate asymptotic
formula for $D_n(t)$ which we now describe.  Introduce another parameter
$s$ and suppose that $n$ and $t$ are related by $n=[2t^{1/2}+s t^{1/6}]$.  Then
as $t\ra\iy$ with $s$ fixed one has
\[ \lim_{t\ra\iy} e^{-t} D_n(t)=F_2(s) \]
where $F_2(s)$ is the distribution function (\ref{F2}).  Using a dePoissonization
lemma due to Johansson~\cite{johansson1}, 
these asymptotics led Baik, Deift and Johansson to the
limiting law
\[ \lim_{N\ra\iy} \textrm{Prob}\left( {\ell_N-2\sqrt{N}\ov N^{1/6}} < s\right)= F_2(s).\]

Since the work of Baik, Deift and Johansson, several groups have
extended this connection
between RSK type combinatorics and the distribution functions of random
matrix theory.
The aforementioned result is equivalent to the determination of
the limiting distribution of the number of
boxes in the first row in the RSK correspondence $\sigma\leftrightarrow
(P,Q)$.
In \cite{bdj2} the same authors show that the limiting distribution of
the number of
boxes in the \textit{second} row  is (when centered and normalized)
distributed as the
\textit{second} largest scaled eigenvalue in GUE \cite{tw1}.
They then conjectured that this correspondence extends to all rows.
This
conjecture was recently proved by A.~Okounkov \cite{okounkov} using
topological methods
and by A.~Borodin, A.~Okounkov and G.~Olshanski \cite{borodin2}
and Johansson~\cite{johansson3} using
analytical methods.

Placing restrictions on the permutations $\sigma$ (that they be fixed
point free and
involutions), Baik and Rains~\cite{baikRains} have shown that the
limiting laws for
the length of the longest increasing/decreasing
subsequence are now the limiting distributions $F_1$ and $F_4$
for the scaled largest eigenvalue in GOE and GSE, see (\ref{F1}) and (\ref{F4}).
Generalizing to signed permutations and colored permutations the present
authors and
Borodin \cite{tw7,borodine1} showed that the
distribution functions of the length of the longest
increasing
subsequence involve the same $F_2$. 

Johansson~\cite{johansson2} showed that the shape
fluctuations of a certain random growth model, again appropriately
scaled, converges in distribution to $F_2$.  (This random growth model is
intimately related to certain randomly growing Young diagrams.)
In subsequent work, Johansson~\cite{johansson3} showed that
the fluctuations in certain random tiling problems (related
to the Artic Circle Theorem) are again described by $F_2$. 
Finally, Johansson~\cite{johansson3}
and the present authors~\cite{tw8} have considered analogous problems
for random \textit{words} and have discovered various random matrix theory
connections.  

\textbf{\large Acknowledgments}
The authors have benefited from conversations with A.~Its and
it is a pleasure to acknowledge this. 
The first author thanks J.~Harnad and P.~Winternitz for
their invitation to speak at the workshop \textit{Integrable Systems:
From Classical to Quantum}.
This work was supported, in part,
by the National Science Foundation through grants DMS--9802122 and
DMS--9732687.

\end{document}